\documentclass[]{spie}  
\newcommand{\bvec}[1]{\mbox{\boldmath $#1$}}

  \usepackage{float} 

\usepackage{amsmath,amsfonts,amssymb}
\usepackage{graphicx}
\usepackage[colorlinks=true, allcolors=blue]{hyperref}

\title{First-principles response simulation of wide-gap CdTe-DSDs for the FOXSI solar sounding rocket experiment}

\author[a*]{Shunsaku Nagasawa}
\author[b, c]{Takahiro Minami}
\author[c, d]{Tadayuki Takahashi}
\author[e]{Shin Watanabe}

\affil[a]{\normalsize Space Sciences Laboratory, University of California, Berkeley, 7 Gauss Way, Berkeley, CA 94720, USA}
\affil[b]{Department of Physics, The University of Tokyo, 7-3-1 Hongo, Bunkyo, Tokyo 113-0033, Japan}
\affil[c]{Kavli Institute for the Physics and Mathematics of the Universe (Kavli IPMU, WPI), The University of Tokyo, 5-1-5 Kashiwanoha, Kashiwa, Chiba 277-8583, Japan}
\affil[d]{International Center for Quantum-field Measurement Systems for Studies of the Universe and Particles (QUP, WPI), KEK, Ibaraki 305-0801, Japan}
\affil[e]{Institute of Space and Astronautical Science, Japan Aerospace Exploration Agency (ISAS/JAXA), 3-1-1 Yoshinodai, Chuo-ku, Sagamihara, Kanagawa 252-5210, Japan}

\authorinfo{*Shunsaku Nagasawa: E-mail: nshunsaku@berkeley.edu}

\begin{document} 
\maketitle

\begin{abstract}
We have developed a wide-gap CdTe double-sided strip detector (CdTe-DSD) for the fourth and fifth flights of the Focusing Optics X-ray Solar Imager sounding rocket experiment (FOXSI-4/FOXSI-5).
This detector features a 30~$\mathrm{\mu}$m strip width and a variable gap width from 30~$\mathrm{\mu}$m to 70~$\mathrm{\mu}$m, enabling position resolution finer than the strip pitch by inducing charge sharing across adjacent strip electrodes and utilizing this shared energy information for position reconstruction.
However, this wide-gap configuration introduces complex detector responses, such as charge loss in the gap regions and electric field distortion near the electrodes, requiring a more advanced modeling approach for interpreting observation results in FOXSI-4/FOXSI-5 and for further optimization of future detector design.
To address these complexities, we have constructed a first-principles simulation framework to model the detector response.
The simulation integrates Geant4-based Monte Carlo simulations of energy deposition, finite element calculations of electric and weighting fields using COMSOL Multiphysics, charge transport modeling incorporating trapping and diffusion, and calculation of the induced charge on the electrodes based on the Shockley-Ramo theorem.
By introducing a surface conductive layer, which causes electric field distortion, the model successfully reproduces the experimentally observed charge loss on the cathode side.
In addition, the model reproduces the distinct charge sharing behavior on the cathode and anode sides.
These results validate the effectiveness of the model in characterizing the wide-gap CdTe-DSD.
\end{abstract}


\section{Introduction}
\label{sec:intro} 
The Focusing Optics X-ray Solar Imager (FOXSI)\cite{krucker2013focusing} is a series of solar-dedicated sounding rocket missions that perform direct focusing imaging spectroscopic observations of solar X-rays using high-resolution Wolter-I optics and fine-pitch focal plane detectors.
Compared to previous indirect imaging missions such as Reuven Ramaty High-energy Solar Spectroscopic Imager (RHESSI)\cite{lin2003reuven, hurford2003rhessi}, FOXSI achieves orders-of-magnitude improvements in sensitivity and imaging dynamic range, enabling the detection of faint X-ray sources from the Sun.
The fourth and fifth flights, FOXSI-4\cite{buitrago2021foxsi, savage2022first} and the upcoming FOXSI-5, aim to perform the first direct focusing observation of medium- and large-scale solar flares ($\geq$ GOES C5 class) in soft and hard X-rays.
These observations are designed to deepen our understanding of flare-related particle acceleration by identifying where particles are energized in the solar corona, clarifying how energetic electrons propagate and lose energy, and elucidating the processes by which accelerated particles escape into interplanetary space.
Addressing these fundamental questions requires focal plane detectors with both high energy resolution and fine position resolution, enabling precise measurements of the spatial and spectral properties of flare-accelerated particles.

To meet these requirements, we developed wide-gap CdTe double-sided strip detectors (CdTe-DSDs) as a hard X-ray focal plane detector in the 4–20~keV range\cite{nagasawa2023wide,nagasawa2025imaging}.
The detectors feature a fine strip pitch with a variable gap structure to enhance charge sharing, enabling sub-strip-pitch position resolution.
We established phenomenological energy and position reconstruction methods for the wide-gap CdTe-DSD by fully utilizing both cathode and anode side signals and the charge sharing information between adjacent strips \cite{nagasawa2025imaging}.
These methods successfully achieved an energy resolution of 0.75 keV (FWHM) at 13.9 keV and a position resolution of 20–50~$\mathrm{\mu m}$, depending on the interaction position within the strip and gap regions.
However, these approaches relied on empirical calibration based on the observed detector response, while the underlying charge transport and loss mechanisms remained to be theoretically understood.
To further optimize detector designs and reconstruction algorithms, and to accurately interpret flare observations in FOXSI-4 and FOXSI-5, it is essential to establish a first-principles understanding of charge sharing and charge loss in wide-gap CdTe-DSDs.

In this paper, we present a comprehensive simulation framework that models the electric field distribution, charge carrier transport, and induced signal formation in wide-gap CdTe-DSDs.
This framework enables a quantitative investigation of how electrode geometry, particularly the wide-gap structure, affects charge sharing and charge collection efficiency.
Section~\ref{sec:det_all} describes the design and key response characteristics of the wide-gap CdTe-DSD.
Section~\ref{sec:sim_all} details the simulation methodology, including electric field calculations, charge transport modeling, and induced signal calculations based on the Shockley–Ramo theorem.
Finally, in Section~\ref{sec:exp_all}, we compare the simulation results with experimental data and discuss the impact of charge loss and charge sharing, with particular focus on the role of surface conductivity in altering the electric field structure and detector response.

\section{Wide-gap CdTe Double-sided Strip Detector}\label{sec:det_all}
\subsection{Detector Configuration}
\begin{figure}[h]
    \centering
    \includegraphics[width=1\linewidth]{ 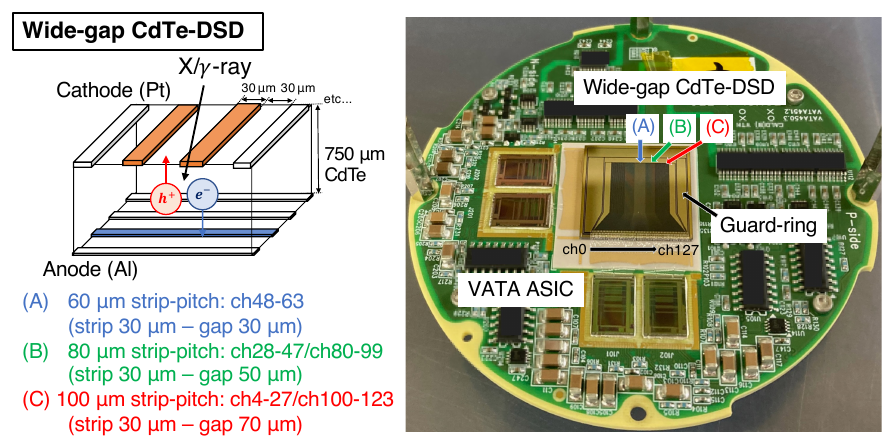}
    \caption{Schematic diagram and picture of the wide-gap CdTe-DSD for FOXSI-4 and FOXSI-5.}
    \label{fig:dsd_geo}
\end{figure}

Figure \ref{fig:dsd_geo} shows the wide-gap CdTe-DSD developed for FOXSI-4 and FOXSI-5. 
The detector has a thickness of 750 $\mathrm{\mu m}$ with a sensitive area of 9.92$\times$9.92 $\mathrm{mm^2}$. 
A total of 128 strip electrodes are orthogonally placed on both the cathode (Pt) and anode (Al) sides.
A unique feature of the detector is its strip layout: while the strip width is fixed at 30 $\mathrm{\mu m}$, the gap between the strips increases from the center toward the edges of the detector, with widths of 30 $\mathrm{\mu m}$, 50 $\mathrm{\mu m}$, and 70 $\mathrm{\mu m}$ in Regions (A), (B), and (C), respectively (see Figure \ref{fig:dsd_geo} and Table \ref{ch_map}).
By adopting this wide-gap structure, charge sharing between adjacent strips is enhanced, enabling the use of this information to improve position resolution\cite{nagasawa2023wide, nagasawa2025imaging}. 
This detector also provides an opportunity to investigate how variations in strip and gap widths affect charge sharing properties, contributing to the further optimization of detector configurations and energy/position reconstruction algorithms for future satellite missions.
Guard-ring electrodes surround the detector to suppress surface leakage current, thereby improving energy resolution~\cite{nakazawa2004improvement}. 
Signals are read from both sides using four VATA451.2 ASICs\cite{watanabe2014si}, each consisting of 64 channels that include a preamplifier, a fast shaper for self-triggering, and a slow shaper for pulse-height measurement (Further details of the readout electronics system can be found in Refs.\citenum{nagasawa2025imaging} and \citenum{nagasawa2024doctor}).
\begin{table}[htb]
  \centering
  \caption{Strip configuration of the wide-gap CdTe-DSD}
  \label{ch_map}
  \begin{tabular}{l|l}
  \hline
  0 ch -- 3 ch & Guard-ring electrodes \\
  4 ch -- 27 ch & (C) 100~$\mathrm{\mu m}$ strip-pitch (strip 30~$\mathrm{\mu m}$ -- gap 70~$\mathrm{\mu m}$)\\
  28 ch -- 47 ch & (B) 80~$\mathrm{\mu m}$ strip-pitch ~~(strip 30~$\mathrm{\mu m}$ -- gap 50~$\mathrm{\mu m}$)\\
  48 ch -- 63 ch & (A) 60~$\mathrm{\mu m}$ strip-pitch ~~(strip 30~$\mathrm{\mu m}$ -- gap 30~$\mathrm{\mu m}$)\\
  80 ch -- 99 ch & (B) 80~$\mathrm{\mu m}$ strip-pitch ~~(strip 30~$\mathrm{\mu m}$ -- gap 50~$\mathrm{\mu m}$)\\
  100 ch -- 123 ch & (C) 100~$\mathrm{\mu m}$ strip-pitch (strip 30~$\mathrm{\mu m}$ -- gap 70~$\mathrm{\mu m}$)\\
  124 ch -- 127 ch & Guard-ring electrodes  \\
  \hline
  \end{tabular}
\end{table}

\subsection{Detector Response}
While widening the gap increases the available information obtained from charge sharing events, it also complicates the detector response and requires consideration of various factors.
In particular, to achieve high energy and position resolution with the CdTe-DSD, it is essential to fully understand the properties of charge sharing and charge loss, so that their effects can be properly utilized and corrected in the reconstruction process.
The following sections describe these key processes in detail.


\subsubsection{Depth of Photon Interaction (DOI) Dependence due to Charge Trapping}
While a pixel-type detector reads signals only from the anode side, the CdTe-DSD independently measures the induced charge on both the anode and cathode sides. 
Therefore, if the energy resolution is dominated by the noise of the readout electronics, the energy resolution is expected to improve by taking the average of both sides of energies.
However, in the case of CdTe, energy reconstruction cannot be performed by simple averaging because the hole mobility is particularly small, causing charge trapping.
To mitigate this effect, the CdTe-DSD is operated with X-rays incident from the cathode side.
Nevertheless, when high-energy X-rays ($E \gtrsim 15$ keV) interact within the detector, the interaction depth increases (closer to the anode side), and the total induced charge becomes dependent on the distance between the interaction point and the electrodes due to hole trapping.
This depth dependence leads to a low-energy tail in the spectrum and degrades the energy resolution.

\subsubsection{Charge Sharing Characteristics between Adjacent Strips}
For the third flight, FOXSI-3, we developed a CdTe-DSD with a fine strip pitch of 60 $\mathrm{\mu m}$ (50 $\mathrm{\mu m}$ strip and 10 $\mathrm{\mu m}$ gap width) to achieve high position resolution\cite{furukawa2019development, furukawa2020imaging}.
In this fine-pitch configuration, the charge cloud generated by an incident photon often spreads across multiple strips. Depending on the interaction position within the detector, signals may appear not only in a single channel (single-strip events) but also in multiple adjacent channels (charge sharing events, or double-strip events when detected in two channels).
Understanding these charge sharing characteristics is essential for accurately reconstructing the energy and interaction position of incident X-rays.

\begin{figure}[h]
    \centering
    \includegraphics[width=1\linewidth]{ 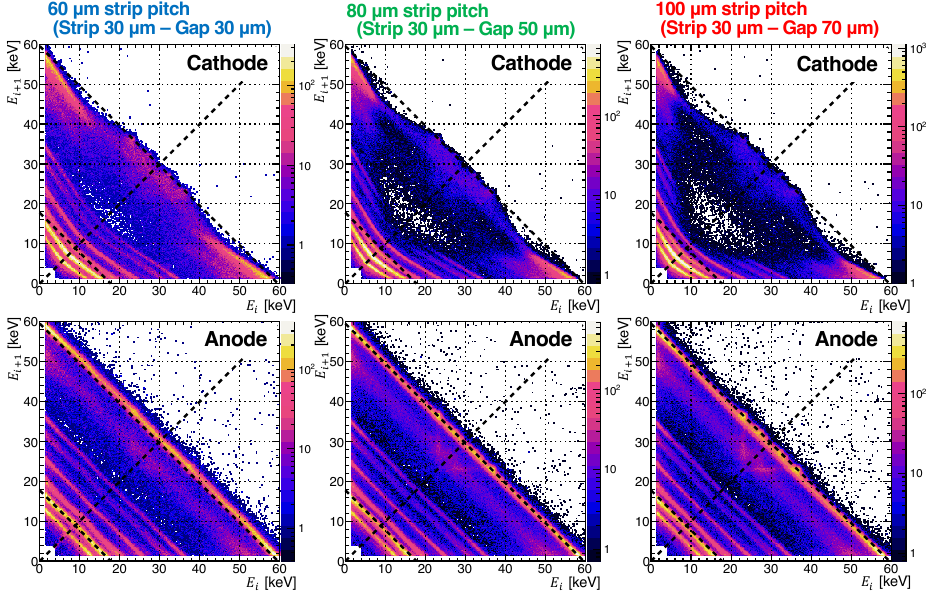}
    \caption{Relationships between the energies detected on adjacent strips for (Upper) the FOXSI-3 CdTe-DSD and (Lower) the wide-gap CdTe-DSD for double-strip events of ${}^{241}$Am. The four points in the 60 keV peak line are the X-ray fluorescence peaks of Cd and Te. The bias voltage is 200 V and the operating temperature is -20${}^\circ$C.}
    \label{fig:spect_diff}
\end{figure}

In particular, in the wide-gap CdTe-DSD developed for FOXSI-4 and FOXSI-5, widening the gaps between strips introduces additional charge loss, which degrades the spectral performance.
Figure \ref{fig:spect_diff} shows the energy correlation between adjacent strips for double-strip events for the wide-gap CdTe-DSD.
For the FOXSI-3 CdTe-DSD, the sum of the energies detected on adjacent strips ($E_{\mathrm{sum}} = E_i + E_{i+1}$, where $E_i$ and $E_{i+1}$ are the energies on the $i$-th and $(i+1)$-th strips, respectively) remains nearly constant, indicating that a simple summation provides a reasonable estimate of the incident photon energy (see also Figure 6 in Ref.\citenum{nagasawa2025imaging}).
In contrast, for the wide-gap CdTe-DSD, the sum deviates from this constant value and becomes smaller, especially on the cathode side and in the regions with the widest gaps.
This charge loss is particularly evident for events where the energies of adjacent strips are comparable ($E_i \simeq E_{i+1}$), which occurs when the photon interacts near the center of the gap.
Therefore, in the wide-gap CdTe-DSD, simple energy summation of adjacent strips is not sufficient for accurate energy reconstruction.

\section{Simulation Framework for Response Modeling} \label{sec:sim_all}
\subsection{Simulation Framework Overview} 
The detection principle of semiconductor detectors such as the wide-gap CdTe-DSD involves several fundamental processes.
First, the energy of an incident X-ray is transferred to the semiconductor via the photoelectric effect, generating photoelectrons and secondary particles.
These energetic charged particles create electron-hole pairs along their trajectories as they travel through the semiconductor.
For CdTe, the energy required to create one electron-hole pair is 4.43 eV \cite{kolanoski2020particle}, and the total number of pairs generated is proportional to the incident X-ray energy.
The generated electron-hole pairs then drift under the electric field applied by the bias voltage between the electrodes.
As they move, they induce charge on the electrodes, which is converted into a voltage signal by a pulse-shaping analog circuit.
The peak voltage of this signal is proportional to the total induced charge.
Accurately modeling the response of the wide-gap CdTe-DSD requires simulating each of these processes appropriately.

For double-sided strip detectors, simulators for modeling the response of the detector have been developed \cite{odaka2010development, hagino2012imaging}.
These conventional simulations assume a narrow gap width much smaller than the strip width, enabling analytical solutions under a parallel electric field and simplified boundary conditions.
However, in the wide-gap CdTe-DSD, the gap width becomes comparable to the strip width, and the electric field is significantly distorted near the gap regions.
Under these conditions, the assumptions used in conventional models no longer hold, and a new simulation framework that incorporates the complex electrode structure is required.

\begin{figure}[h]
    \centering
    \includegraphics[width=1\linewidth]{ 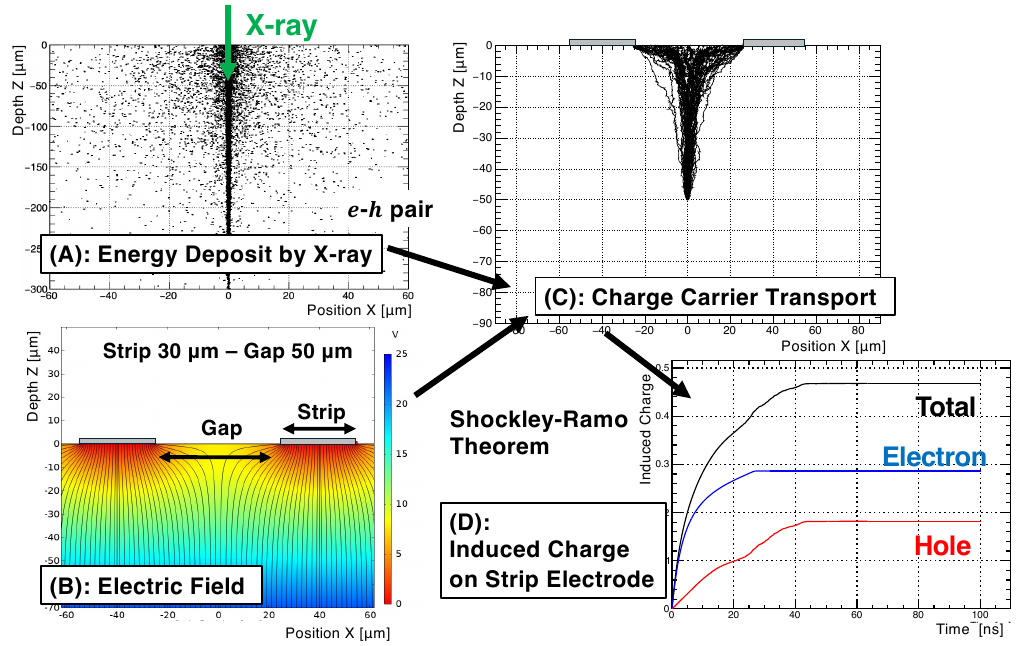}
    \caption{Simulation framework for modeling the response of the wide-gap CdTe-DSD}
    \label{fig:sim_framework}
\end{figure}

A schematic overview of the simulation framework developed in this study is shown in Figure \ref{fig:sim_framework}.
The total induced charge on the strip electrodes is calculated through the following procedure:
\begin{description}
  \item[\rm (A): Energy deposition in the CdTe semiconductor by incident X-ray: Section \ref{sec:geant4_map}]\mbox{}\\
  The interaction and energy deposition of incident X-ray in the CdTe crystal are simulated using the Monte Carlo simulation toolkit Geant4 (GEometry ANd Tracking) \cite{agostinelli2003geant4}.
  The deposited energy is then converted to the corresponding number of electron-hole pairs.
  \item[\rm (B): Calculation of the electric field: Section \ref{sec:elect}]\mbox{}\\
  Due to the complexity of the electrode structure, an analytical solution of the Poisson equation is impractical. Therefore, the electric field distribution is calculated using the Finite Element Method (FEM) implemented in COMSOL Multiphysics~\cite{dickinson2014comsol}.
  \item[\rm (C): Charge carrier transport: Section \ref{sec:cartrans}]\mbox{}\\ 
  Based on the initial positions of the generated electron-hole pairs and the calculated electric field, the trajectories of charge carriers are determined by solving the equations of motion, including the effects of charge diffusion.
  \item[\rm (D): Calculation of the induced charge on the strip electrodes: Section \ref{sec:induced_shockley}]\mbox{}\\
  The induced charge on the strip electrodes is calculated from the charge carrier trajectories using the Shockley–Ramo theorem~\cite{shockley1938currents, ramo1939currents, he2001review}.
\end{description}

\subsection{Energy Deposition in CdTe Semiconductor by Incident X-ray}\label{sec:geant4_map}
To simulate the interactions between incident X-rays and the CdTe semiconductor detector, we conducted Monte Carlo simulations using the Geant4 toolkit (version 11.2.2)~\cite{agostinelli2003geant4}.
Geant4 is a Monte Carlo simulation framework for modeling the interactions of photons and particles as they pass through matter, utilizing a comprehensive database of physical models and experimentally measured cross-sections.
It consists of object-oriented class libraries written in C++, allowing users to flexibly define the material structures, incident particle properties, and interaction processes.
In this study, we employed the ``G4EmLivermorePhysics"  library as the electromagnetic interaction model.
This library specializes in modeling electromagnetic interactions between photons and matter over an energy range from approximately 250 eV to several GeV.
By using this package, we account for physical processes such as the photoelectric effect, Compton scattering, and Rayleigh scattering for photons, as well as ionization and multiple scattering for electrons.

\begin{figure}[h]
    \centering
    \includegraphics[width=1\linewidth]{ 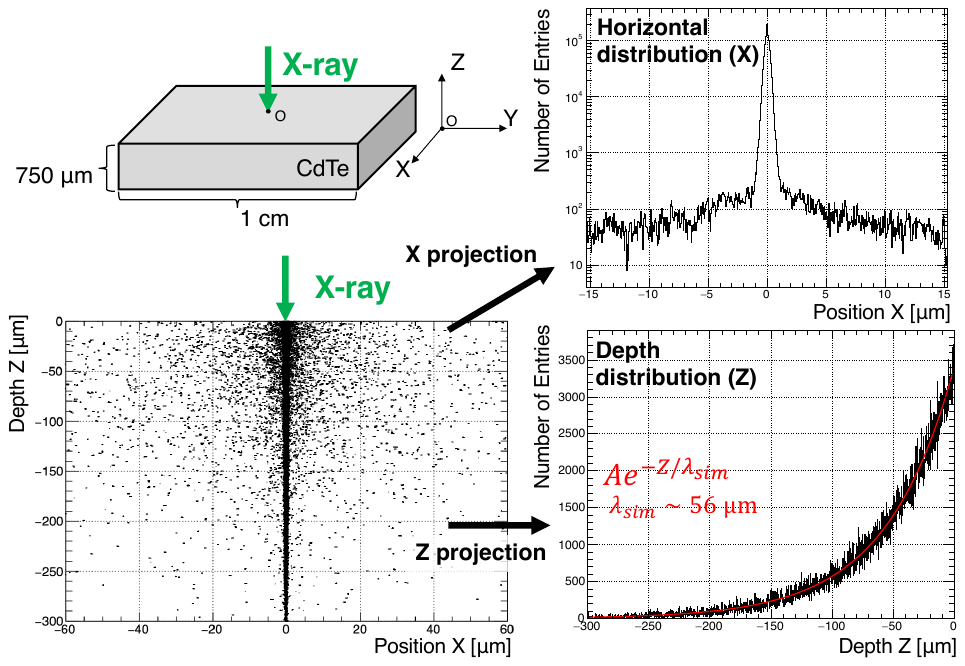}
    \caption{(Upper left) Schematic diagram of the simulation setup. (Lower left) Simulated energy deposition distribution in CdTe by a 17.8 keV X-ray beam. (Upper and lower right) Projected distributions along the horizontal ($X$) and depth ($Z$) directions.}
    \label{fig:sim_geant4}
\end{figure}

We simulated the distribution of energy deposition in CdTe by injecting a monochromatic X-ray beam.
Figure~\ref{fig:sim_geant4} (upper left) shows a schematic diagram of the simulation setup.
A CdTe semiconductor box with a thickness of 750 $\mathrm{\mu m}$ is placed with its upper surface defined as $Z = 0$, and the X-ray beam is injected along the negative $Z$-axis.
The energy of the X-ray beam is set to 17.8 keV, corresponding to the energy peak of ${}^{241}$Am.
In total, $10^6$ X-ray photons were injected to simulate the energy deposition distribution.

Figure~\ref{fig:sim_geant4} (lower left) shows a scatter plot of the energy deposition positions.
The upper and lower right panels show the projected distributions in the horizontal ($X$) and depth ($Z$) directions, respectively.
In the horizontal distribution, since the typical travel length of 17.8 keV photoelectrons is approximately 2 $\mathrm{\mu m}$~\cite{navas2024review},
the energy deposition is confined within $\pm$2 $\mathrm{\mu m}$ and does not spread in the horizontal direction.
A few Rayleigh scattering events are observed, extending up to about 20 $\mathrm{\mu m}$.
In the depth distribution, the interaction positions follow an exponential distribution with a characteristic length of approximately 56 $\mathrm{\mu m}$,
consistent with the attenuation length of 17.8 keV X-rays in CdTe ($\sim$59 $\mathrm{\mu m}$)~\cite{henke1993x}.
In summary, for 17.8 keV X-rays, the energy deposition profile exhibits an exponential attenuation along the incident beam direction, with only a slight lateral spread.

\subsection{Calculation of Electric Field using Finite Element Method} \label{sec:elect}

\begin{figure}[hp]
    \centering
    \includegraphics[width=1\linewidth]{ 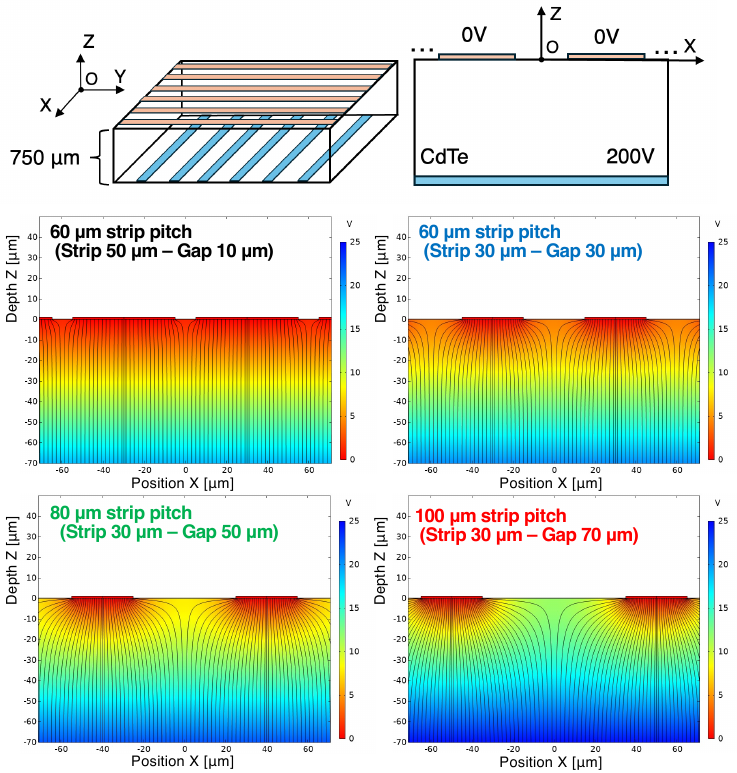}
    \caption{(Upper) Schematic diagram of the simulation geometry. For simplicity, calculations are performed in two dimensions ($X$-$Z$ plane). (Lower) Electric potentials $V(\bvec{x})$ and electric field lines of the FOXSI-3 CdTe-DSD and wide-gap CdTe-DSD for each strip/gap width region, calculated by applying 0 V to the upper electrodes ($Z = 0$) and 200 V to the lower electrode ($Z = -750~\mathrm{\mu m}$).}
    \label{fig:sim_efield}
\end{figure}

The Finite Element Method (FEM) is a computational technique for obtaining approximate numerical solutions of partial differential equations in domains with complex geometries and boundary conditions.
In this study, we calculated the electric field distributions of the wide-gap CdTe-DSD in each strip and gap region using the Electric Currents module of COMSOL Multiphysics~\cite{dickinson2014comsol}(\url{https://www.comsol.com}), a general-purpose FEM-based simulation platform for a wide range of physics domains, including electromagnetics.
As shown in Figure~\ref{fig:sim_efield} (upper), the calculations are performed in two dimensions (in the $X$-$Z$ plane) for simplicity.
The simulation domain is discretized into small elements, and an adaptive mesh is automatically generated in COMSOL according to the physics and geometry.
The mesh is refined near the electrodes, where the electric field gradient is steep, and coarser elsewhere to optimize computational efficiency.

Figure~\ref{fig:sim_efield} shows the electric potentials $V(\bvec{x})$ of the wide-gap CdTe-DSD for each strip/gap width region, calculated by applying 0 V to the upper electrodes ($Z = 0$) and 200 V to the lower electrode ($Z = -750~\mathrm{\mu m}$).
For comparison, the electric potential of the FOXSI-3 CdTe-DSD is also shown.
In the FOXSI-3 CdTe-DSD, the electric potential distribution is nearly parallel due to its narrow gap structure.
In contrast, the wide-gap CdTe-DSD exhibits significant distortion of the electric field near the surface, and this distortion becomes more pronounced as the gap width increases.
Since most X-rays in the FOXSI energy range ($<20$ keV) interact and deposit energy near the surface of the CdTe, this field distortion is expected to affect the transport of charge carriers, particularly on the cathode side.

\subsection{Charge Carrier Transport} \label{sec:cartrans}
Based on the calculated electric field, we model the transport of electrons and holes generated by incident X-rays.
The effects considered in this simulation include charge drift in the electric field, charge diffusion, and charge trapping.
\subsubsection{Charge Drift}
Charge carriers generated by X-rays drift in the electric field $E(\bvec{x})$ with a velocity given by $\bvec{v}(\bvec{x}) = \pm \mu \bvec{E}(\bvec{x})$, where $\mu$ is the carrier mobility.
The sign corresponds to the charge $q$ of the carrier.
In CdTe, the mobilities of electrons and holes are approximately 1050 $\mathrm{cm^2/V \cdot s}$ and 90 $\mathrm{cm^2/V \cdot s}$, respectively~\cite{kolanoski2020particle}.
For example, when an X-ray interacts at a depth of $Z = -59~\mathrm{\mu m}$, which corresponds to the attenuation length of a 17.8 keV X-ray, the distance to the anode side is 691 $\mathrm{\mu m}$ and to the cathode side is 59 $\mathrm{\mu m}$.
Under a uniform bias voltage of 200 V, the estimated drift times of electrons and holes to reach the anode and cathode electrodes are both approximately 25 ns.

\subsubsection{Charge Diffusion}
The charge cloud generated by X-rays spreads due to thermal motion and diffuses into the surrounding material.
Here, we consider a charge cloud with a time-dependent distribution $\rho(\bvec{x}, t)$ in the absence of external fields.
The time evolution of the charge cloud follows the diffusion equation (Fick’s second law):
\begin{equation}
  \frac{\partial \rho (\bvec{x}, t)}{\partial t} = D \nabla^2 \rho (\bvec{x}, t)
  \label{eq_diff}
\end{equation}
where $D$ is the diffusion coefficient, assumed to be constant and isotropic.
In principle, the distributions of electrons and holes can be obtained by solving the drift and two-dimensional diffusion equations, incorporating the electric field calculated in Section~\ref{sec:elect}.
However, directly solving these advection-diffusion equations requires high computational resources and may lead to numerical instability.
Therefore, in this study, we approximate the time evolution of the carrier distribution by tracking a collection of ``test particles" and model the charge diffusion as a random walk.
Then, the time evolution for each test particle is randomly determined by:
\begin{eqnarray}
x(t + \Delta t) &=& x(t) + \cos{\theta} \Delta r \\
z(t + \Delta t) &=& z(t) + \sin{\theta} \Delta r
\end{eqnarray}
where, $\Delta r$ is a step size for a time step $\Delta t$, and $\theta$ is a random variable uniformly distributed from 0 to $2\pi$.
The step size $\Delta r$ is related to the diffusion coefficient $D$ by $\Delta r = \sqrt{4 D \Delta t}$.
The diffusion coefficient $D$ for thermal diffusion is given by the Einstein relation, $D = \mu k_B T / e$, where $k_B$ is the Boltzmann constant and $T$ is the absolute temperature.
At $T = -20^\circ \mathrm{C}$, the diffusion coefficient is approximately 23~$\mathrm{cm^2/s}$ for electrons and 2.0~$\mathrm{cm^2/s}$ for holes.
Since the FOXSI observations are limited to photon energies below 20~keV, Coulomb repulsion effects during drift are negligible and thus not included in the simulation\cite{ballester2024charge}.

\subsubsection{Charge Trapping}
In the charge diffusion calculation described above, the number of carriers is conserved.
In practice, however, the number of charge carriers decreases over time due to charge trapping, which results from the recombination of electrons and holes before they reach the anode or cathode electrodes.
The decay of the carrier population follows an exponential law, given by $N_e(t) = N_0 \exp(-t/\tau_e)$ and $N_h(t) = N_0 \exp(-t/\tau_h)$ for electrons and holes, respectively.
Here, $N_0$ is the initial number of carriers in the charge cloud at $t = 0$, and $\tau_e$ and $\tau_h$ are the carrier lifetimes for electrons and holes, respectively.
For CdTe, these lifetimes are approximately 3 $\mathrm{\mu s}$ for electrons and 1 $\mathrm{\mu s}$ for holes~\cite{grimm2020signal}.

\subsection{Induced Charge on Strip Electrode}\label{sec:induced_shockley}
To calculate the charge induced on the electrodes by the motion of charge carriers, we apply the Shockley–Ramo theorem\cite{shockley1938currents, ramo1939currents, he2001review}, which provides an efficient method to calculate induced charge without solving the full set of Maxwell’s equations.
In this theorem, the mathematical concept of ``weighting potential~$\phi_{w}$" is introduced. 
When a charge carrier moves within the detector, it induces a signal on the electrodes.
According to the Shockley–Ramo theorem, the induced charge $Q$ on an electrode (either on the cathode or anode side) is calculated from the carrier trajectory and the corresponding weighting field.
Specifically, for a carrier with charge density $q(\vec{r})$ moving from its initial position $\vec{r_0}$ to its final position $\vec{r_f}$, the induced charge is given by:
\begin{align}
Q=\int^{\vec{r_f}}_{\vec{r_0}} q(\vec{r'})\vec{E_w}(\vec{r'})\cdot d\vec{r'},
\label{chargeeff}
\end{align} 
where $\vec{E_w} = -\nabla \phi_w$ is the weighting field.
The trajectory of the charge carriers is determined by the actual electric field $\vec{E}$.
To calculate the induced charge accurately using the Shockley–Ramo theorem, it is essential to obtain the weighting potential $\phi_w$ corresponding to the electrode configuration.
The weighting potential $\phi_w$ is calculated by solving Laplace’s equation, $\nabla^2 \phi_w(\bvec{x}) = 0$, with boundary conditions such that the electrode of interest is set to unit potential ($\phi_w = 1$) and all other electrodes are grounded ($\phi_w = 0$).
Since analytical solutions are impractical for the complex electrode structure of the wide-gap CdTe-DSD, we calculate the weighting potential numerically using COMSOL Multiphysics.
The calculation procedure is the same as that used for the electric field in Section~\ref{sec:elect}.

\begin{figure}[hp]
    \centering
    \includegraphics[width=0.9\linewidth]{ 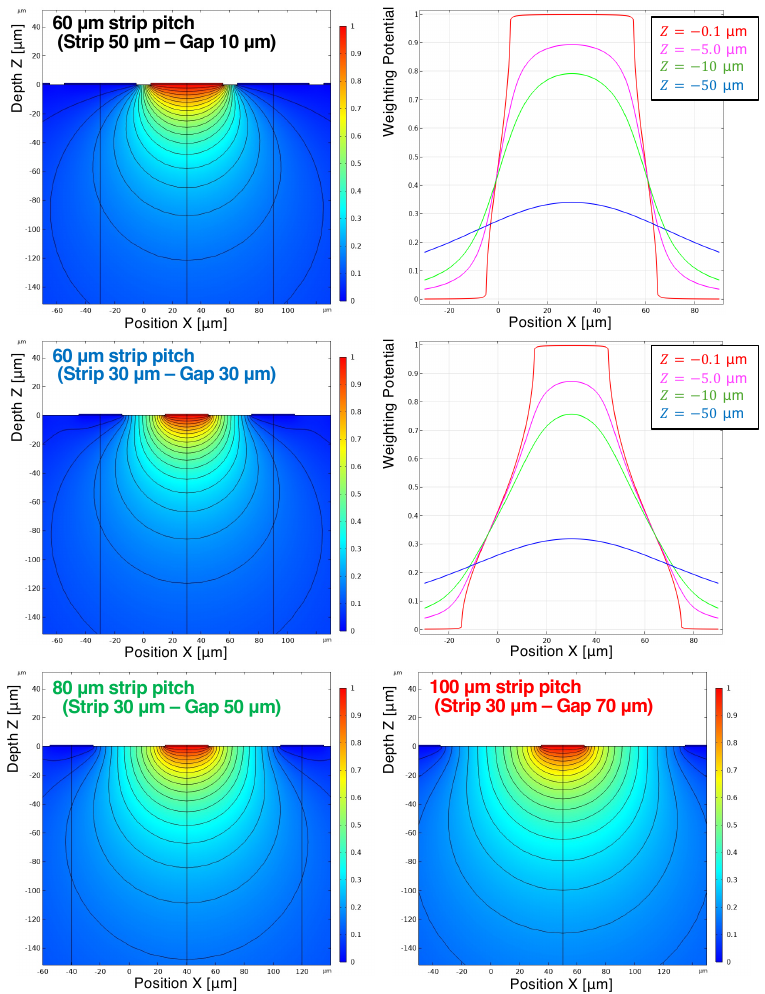}
    \caption{Weighting potentials $\phi_w(\bvec{x})$ of the FOXSI-3 CdTe-DSD and the wide-gap CdTe-DSD for each strip and gap region. Slices along the $X$-axis at Z = -0.1, -5.0, -20, and -50 $\mathrm{\mu m}$ are also shown for the FOXSI-3 CdTe-DSD and the 60~$\mathrm{\mu m}$ strip-pitch region of the wide-gap CdTe-DSD.}
    \label{fig:sim_weight}
\end{figure}

Figure~\ref{fig:sim_weight} shows the weighting potentials $\phi_w(\bvec{x})$ of the wide-gap CdTe-DSD for each strip and gap region.
For comparison, the weighting potential of the FOXSI-3 CdTe-DSD is also shown.
The weighting potential exhibits a steep gradient near the readout electrodes.
Since the induced charge is determined by the difference in weighting potential between the initial and final positions of the carrier, the signal is primarily generated by the carrier motion near the readout electrode.
As a result, the motion of holes and electrons mainly contributes to the signals on the cathode and anode sides, respectively (known as the ``small-pixel effect").
In addition, the weighting potential in the gap region is higher for the wide-gap detector compared with the FOXSI-3 CdTe-DSD.
Consequently, a considerable portion of the induced signal appears not only on the strip directly beneath the interaction point but also on adjacent strips.

\section{Simulation of Carrier Transport and Signal Generation}\label{sec:exp_all}
\subsection{Time Evolution of Carrier Distribution and Induced Charge}\label{sec:time_evol_signal}
The induced charge on the strip electrodes is simulated based on the Shockley–Ramo theorem, using the calculated electric field \( E(\bvec{x}) \) and weighting potential \( \phi_w(\bvec{x}) \).
By incorporating the effects of charge drift, diffusion, and trapping described in Section~\ref{sec:cartrans}, we compute the time evolution of the electron and hole distributions using $2\times10^5$ test particles (indexed by $j$).
The motion of each test particle is determined by the following equations:
\begin{eqnarray}
    x_{e/h}^j(t + \Delta t) &=& x_{e/h}^j(t) + \cos{\theta} \Delta r  \pm \mu_{e/h} E_x( x_{e/h}^j(t), z_{e/h}^j(t) ) \Delta t \\
    z_{e/h}^j(t + \Delta t) &=& z_{e/h}^j(t) + \sin{\theta} \Delta r  \pm \mu_{e/h} E_z( x_{e/h}^j(t), z_{e/h}^j(t) ) \Delta t
\end{eqnarray}
where $\Delta r$ represents the diffusion step, and $\theta$ is a random angle uniformly distributed from 0 to 2$\pi$.
The induced charge $Q$ on a strip electrode is calculated from the change in weighting potential along the carrier trajectory as:
\begin{eqnarray}
  \Delta Q_{e/h}^j &=&
  \pm q \exp\left(-\frac{t}{\tau_{e/h}}\right)\left[ \phi_w(x_{e/h}^j(t), z_{e/h}^j(t)) - \phi_w(x_{e/h}^j(t + \Delta t), z_{e/h}^j(t + \Delta t)) \right] \\
  Q &=& \sum\limits_{\bvec{x} = \bvec{x_i}}^{\bvec{x} = \bvec{x_f}} \sum\limits_{j} \Delta Q_e^j + \sum\limits_{\bvec{x} = \bvec{x_i}}^{\bvec{x} = \bvec{x_f}} \sum\limits_{j} \Delta Q_h^j
\end{eqnarray}
where the exponential term accounts for charge trapping.
The calculation for each particle terminates when it reaches $Z_f = 0~\mathrm{\mu m}$ (cathode surface) or $Z_f = -750~\mathrm{\mu m}$ (anode surface).
Carriers arriving within an electrode gap are lost through surface recombination before reaching the electrode.

Figure~\ref{fig:sim_diffcharge} shows the time evolution of the electron and hole distributions and the induced charge on the cathode electrode.
The initial depth of the carrier generation is set at $Z_i = -60~\mathrm{\mu m}$, corresponding to the attenuation length of 17.8 keV X-rays.
The initial horizontal position is $X_i = 0~\mathrm{\mu m}$, located beneath the center of the cathode-side gap.
When carriers are generated beneath the center of a strip, electrons and holes drift directly toward the anode and cathode electrodes, respectively.
The combined contributions from electrons and holes induce a charge on the strip nearly equal to the total generated charge.
In contrast, when carriers are generated beneath the center of cathode-side gap, electrons still drift directly toward the anode, but holes spread laterally during transport and are partially collected by adjacent electrodes, resulting in charge sharing.

In this case, the hole contribution to the strip signal is reduced, leading to an induced charge on the strip that is approximately half of the total generated charge.
Additionally, the electron distribution exhibits larger diffusion during transport, since the diffusion coefficient is proportional to mobility, $D \propto \mu$.

\begin{figure}[hp]
    \centering
    \includegraphics[width=1\linewidth]{ 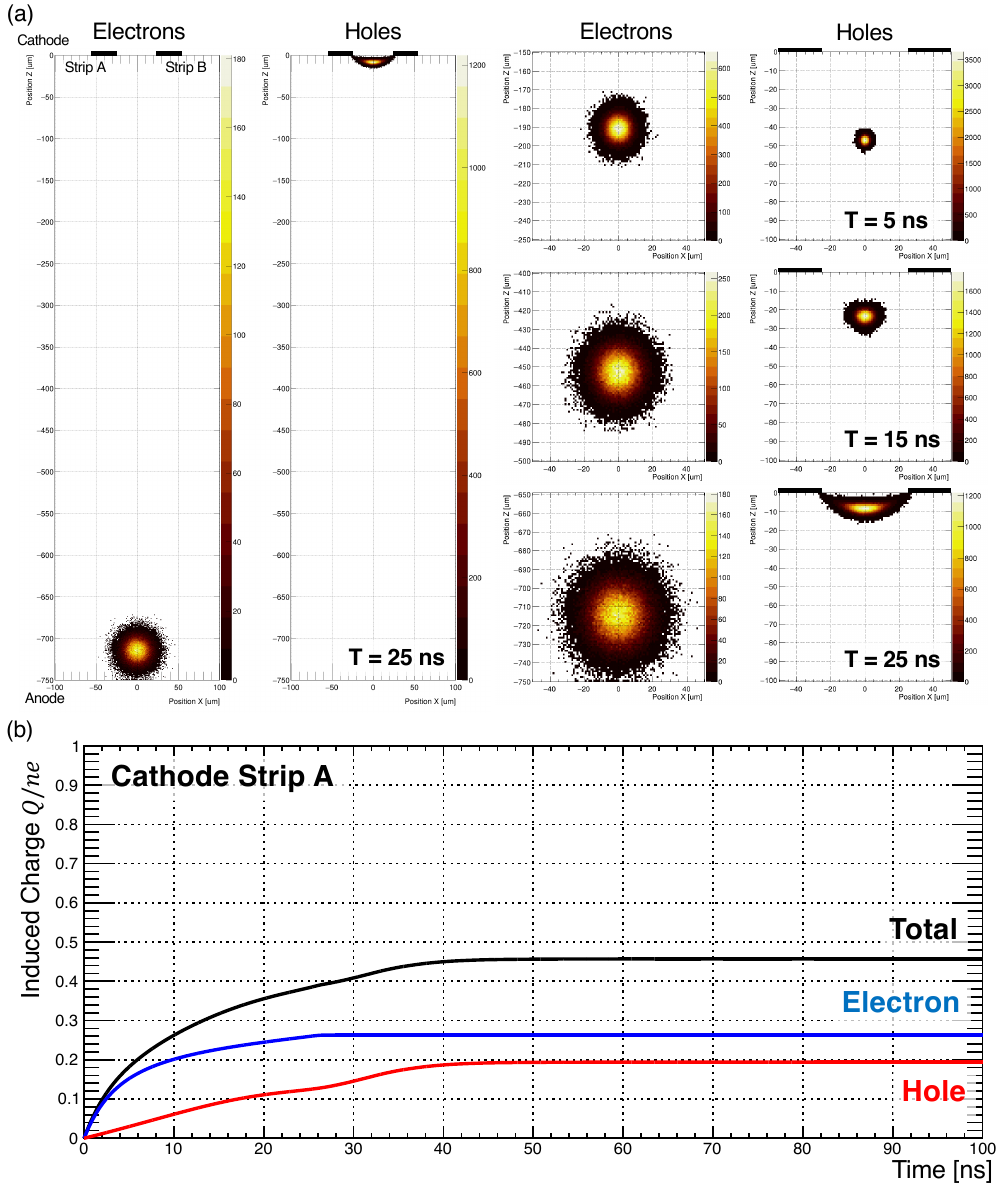}
    \caption{Time evolution of the motion of electrons and holes and the total induced charge on the cathode electrode, for carriers generated at the gap center at $X = 0~\mathrm{\mu m}$ and $Z = -60~\mathrm{\mu m}$. The cathode is located at $Z = 0$ and the anode at $Z = -750~\mathrm{\mu m}$. The induced charge is normalized to the total generated charge}
    \label{fig:sim_diff_movie}
\end{figure}

\subsection{Position Dependence of Charge Collection Efficiency}\label{sec:pos_cce_map}
Following the procedure described in Section~\ref{sec:time_evol_signal}, we calculate the induced charge $Q$ on a strip electrode for any given initial carrier generation position ($X, Z$).
To investigate how the induced charge varies with the interaction position, we define the charge collection efficiency (CCE), as $\eta(X_i, Z_i) = Q(X_i, Z_i) / n e$,
where $Q(X_i, Z_i)$ is the induced charge collected at the electrodes and $ne$ is the total charge generated by the incident X-rays. By definition, $\eta = 1$ when there is no charge loss.

\begin{figure}[h]
    \centering
    \includegraphics[width=0.85\linewidth]{ 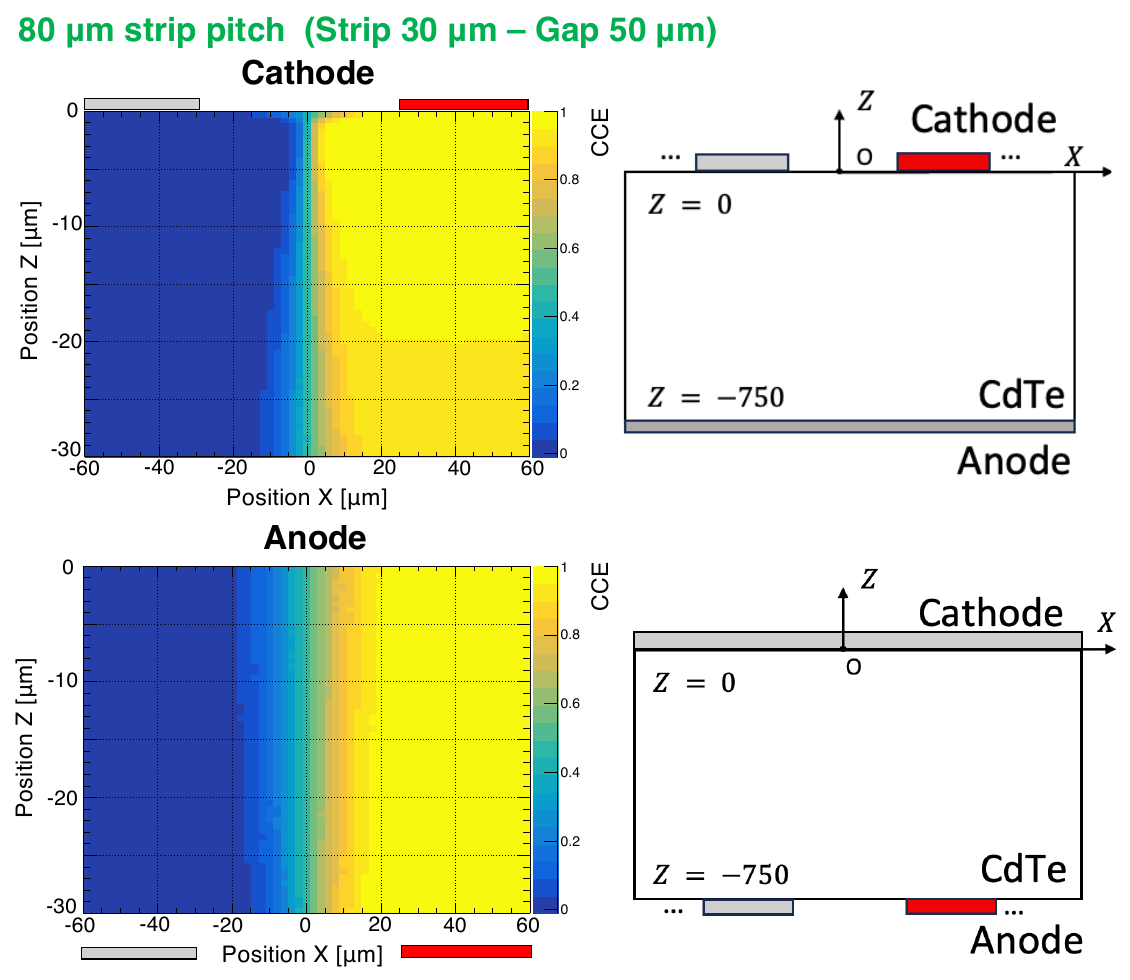}
    \caption{Charge collection efficiency $\eta (X_i, Z_i)$ depends on the initial interaction position for the cathode and anode side.}
    \label{fig:sim_cce}
\end{figure}

First, we calculated a CCE map for both the cathode and anode electrodes by varying the initial carrier generation position in the region with an 80~$\mathrm{\mu m}$ strip pitch.
Figure~\ref{fig:sim_cce} shows the detector geometry and the calculated CCE distributions.
In this configuration, the center of the target strip is at $X = 40~\mathrm{\mu m}$, and the center of the gap is at $X = 0~\mathrm{\mu m}$.
As the interaction depth increases, some charge is still collected even when the initial position is offset from the electrode, due to the effect of diffusion.
This diffusion effect is more pronounced on the anode side, which is located deeper within the detector.
In addition, due to the lower mobility of holes, the induced charge on the cathode side decreases as the interaction depth increases.

\subsection{Comparison with Experimental Results}\label{sec:sim_comp_exp}
Using the CCE map calculated in Section~\ref{sec:pos_cce_map}, we simulated the detected energies on two adjacent strips for various incident X-ray positions $X_i$, and compared the results with experimental data.
For an incident X-ray with energy $E_{\mathrm{in}}$ injected from a horizontal position $X$, the energy deposition $E_{\mathrm{dep}}(X_i, Z_i)$ at each position in the CdTe crystal is calculated using Geant4, as described in Section~\ref{sec:geant4_map}.
The number of generated electron-hole pairs at each point ($X_i, Z_i$) is then obtained by dividing the deposited energy by the ionization energy in CdTe ($\epsilon \sim 4.43$ eV for CdTe\cite{kolanoski2020particle}).
By summing the charge collection efficiency map $\eta(X_i, Z_i)$, weighted by the distribution of generated carriers, the detectable energy $E_{\mathrm{strip}}(X)$ on each strip for an incident X-ray at position $X$ is calculated as:
\begin{equation}
  E_{strip}(X) = E_{in} \frac{\sum_{i} \eta(X_i, Z_i)E_{dep}(X_i,Z_i)/\epsilon}{\sum_{i}E_{dep}(X_i,Z_i)/\epsilon}
\end{equation}

\begin{figure}[h]
    \centering
    \includegraphics[width=0.85\linewidth]{ 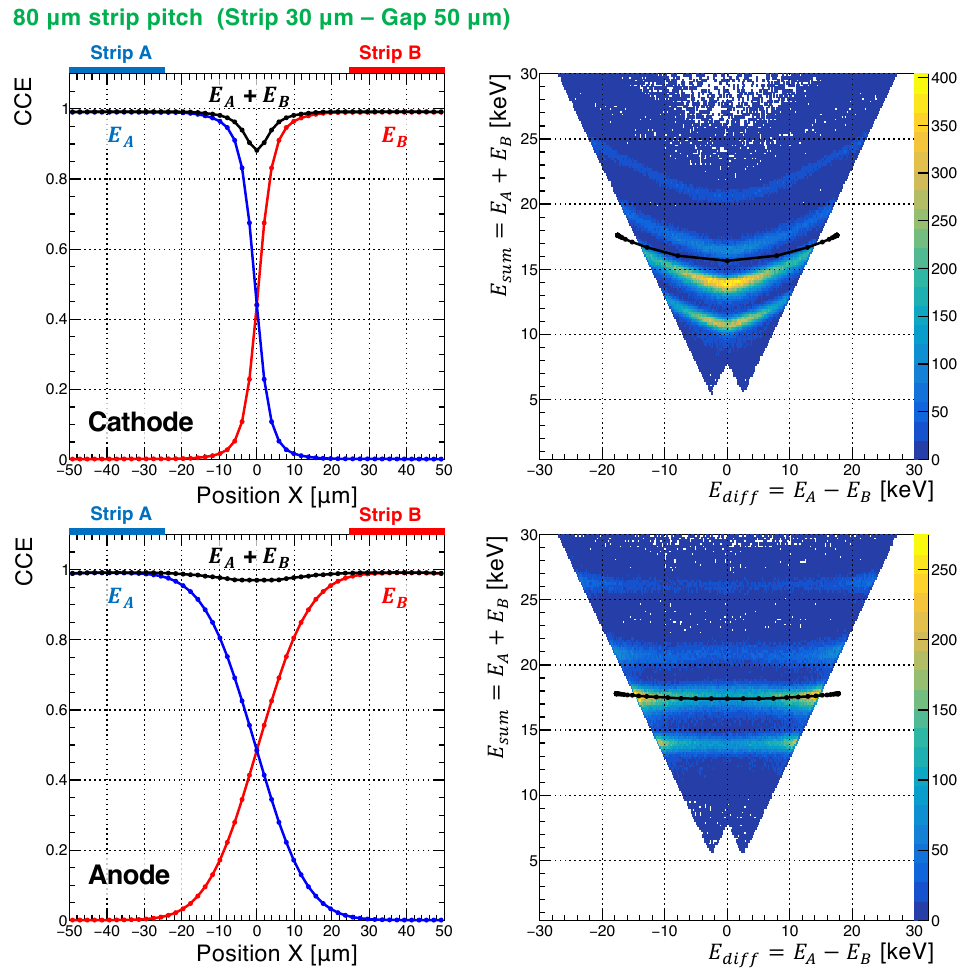}
    \caption{(Left) Simulation result of the detected energy in two adjacent strip A and strip B s a function of the incident X-ray position. (Right) Relationship between the sum and difference of the detected energies on adjacent strips (image: experimental results, black dots: simulation)}
    \label{fig:sim_nocharge}
\end{figure}

The simulation results for 17.8 keV X-rays are shown in Figure~\ref{fig:sim_nocharge} (left).
As the incident position $X$ moves from the strip A toward the strip B, the shared energy varies continuously on both the cathode and anode sides.
This variation is more gradual on the anode side, reflecting the higher probability of charge sharing events in this region.
For comparison with experimental results, Figure~\ref{fig:sim_nocharge} (right) shows the relationship between the sum and difference of the energies detected on adjacent strips, defined as $E_{\mathrm{sum}} = E_{A} + E_{B}$ and $E_{\mathrm{diff}} = E_{A} - E_{B}$.
The simulation results for the anode side reproduce the experimental measurements reasonably well, showing little charge loss.
In contrast, on the cathode side, the simulation underestimates the charge loss compared to the experimental data.
For a 17.8 keV incident X-ray, the simulated charge loss at the gap center relative to the strip center is approximately 10\% (see also Figure~\ref{fig:sim_diff_movie}), which is not sufficient to explain the experimentally observed maximum loss of around 20\%.

\subsection{Effects of Surface Conductive layer on Charge Loss} \label{sec:surface}
In the current simulation setup, the CdTe crystal is modeled with a bulk conductivity of $1 \times 10^{-7}\ \mathrm{S/m}$, corresponding to a typical resistivity of $10^{9}\ \Omega\cdot\mathrm{cm}$\cite{kolanoski2020particle}.
The surface conductivity is initially set equal to the bulk conductivity, and the surface behaves as an intrinsic extension of the bulk.
Under these conditions, the electric field lines curve smoothly toward the strip electrodes, including in the gap regions, so that most of the charge carriers are guided directly to the electrodes before reaching the surface.
This configuration explains the small charge loss observed in the simulation results, as shown in Section~\ref{sec:sim_comp_exp}.

However, actual detectors often exhibit surface layers with higher conductivity due to fabrication effects, the formation of Te-rich layers or surface oxidation.
For example, Ref.\citenum{bolotnikov1999charge} demonstrated that a low-resistivity surface layer on CZT significantly alters the electric field in the inter-pixel regions, causing charge loss when field lines terminate on the surface instead of being directed straight to the electrodes.
Furthermore, Ref.\citenum{bolotnikov2002effects} supported the existence of such a surface layer, reporting that a low-resistivity layer formed during the fabrication process exhibits electrical properties that differ considerably from those of the bulk material and is responsible for the high surface leakage current as well as the degradation of charge collection efficiency in pixel detectors.
In CdTe detectors, a similar low-resistivity surface layer likely also exists and could produce comparable effects.

\begin{figure}[h]
    \centering
    \includegraphics[width=1\linewidth]{ 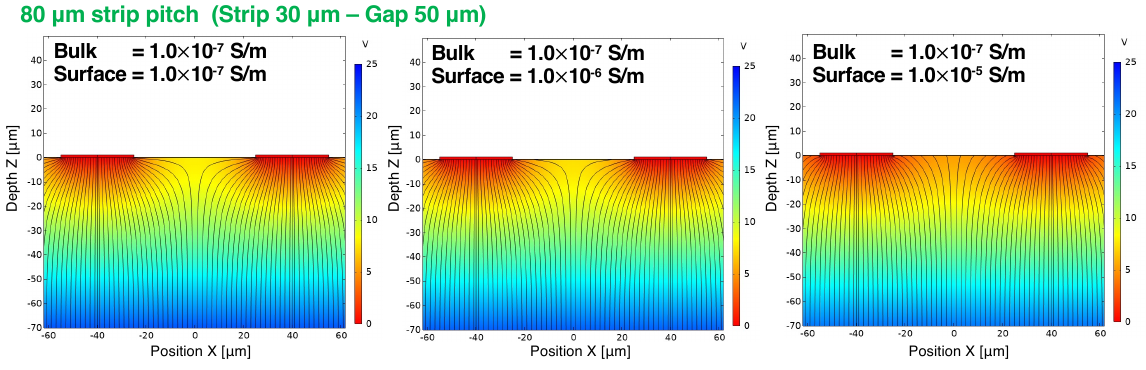}
    \caption{Electric potentials $V{(\bvec{x})}$ of the wide-gap CdTe-DSD in 80 $\mathrm{\mu m}$ strip pitch region for different surface conductivity.}
    \label{fig:sim_diff_surface_ef}
\end{figure}

To examine how the surface conductivity affects the electric field structure and charge collection, we varied the surface conductivity and simulated its influence on the electric field and induced charge.
Figure~\ref{fig:sim_diff_surface_ef} shows the simulated electric potential and field-line distributions for an 80~$\mathrm{\mu}$m strip pitch, where the bulk conductivity is fixed at $1 \times 10^{-7}\ \mathrm{S/m}$ and the surface conductivity is varied across $1 \times 10^{-7}$, $1 \times 10^{-6}$, and $1 \times 10^{-5}\ \mathrm{S/m}$.
As the surface conductivity increases, the electric field beneath the gap becomes more vertically oriented, meaning that the field lines are more likely to terminate on the surface before reaching the electrodes.
This causes charge carriers drifting near the surface to be collected by the surface layer rather than by the electrodes, resulting in charge loss.

\begin{figure}[hp]
    \centering
    \includegraphics[width=0.85\linewidth]{ 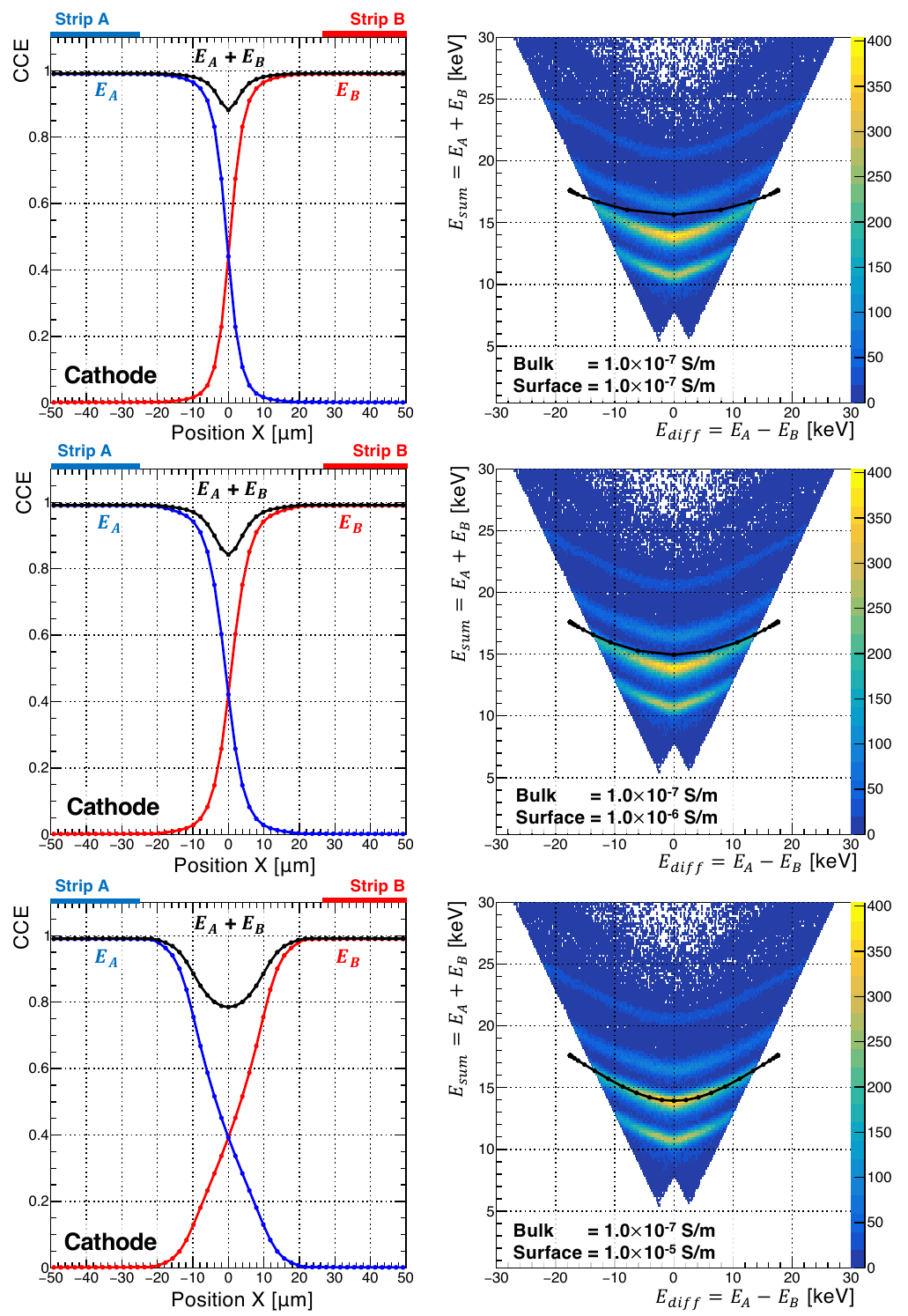}
    \caption{Effects of surface conductivity on the cathode side for 80 $\mathrm{\mu m}$ strip pitch region. (Left): Simulation result of the detected energy (CCE) in two adjacent strips A and B as a function of the incident X-ray position. (Right): Relationship between the sum and difference of the detected energies in adjacent strips (image: experimental results, black dots: simulation).}
    \label{fig:sim_diffcharge}
\end{figure}

Based on these electric field distributions, we calculated the induced charge on the strip electrodes on the cathode side, as shown in Figure~\ref{fig:sim_diffcharge}.
As the surface conductivity increases, the charge loss in the gap region gradually increases.
When the surface conductivity becomes two orders of magnitude higher than that of the bulk, at $1 \times 10^{-5}\ \mathrm{S/m}$, the simulated charge loss reaches approximately 20\%, consistent with the experimental results for 17.8 keV X-rays.
These results suggest that a conductive surface layer may modify the electric field distribution near the surface and contributes to the observed charge loss.


\subsection{Effects of Gap Width on Charge Loss}\label{sec:sim_charge_loss}
As discussed in Section~\ref{sec:surface}, the charge loss observed on the cathode side for an 80~$\mathrm{\mu m}$ strip pitch region (30~$\mathrm{\mu m}$ strip width and 50~$\mathrm{\mu m}$ gap) can be reproduced by introducing a surface conductive layer of $1 \times 10^{-5}\ \mathrm{S/m}$, which is two orders of magnitude higher than the bulk conductivity.
To investigate the dependence of charge loss on gap width, additional simulations were performed for two other strip-pitch configurations: a 60~$\mathrm{\mu m}$ strip pitch (30~$\mathrm{\mu m}$ strip width and 30~$\mathrm{\mu m}$ gap) and a 100~$\mathrm{\mu m}$ strip pitch (30~$\mathrm{\mu m}$ strip width and 70~$\mathrm{\mu m}$ gap). 

\begin{figure}[hp]
    \centering
    \includegraphics[width=0.85\linewidth]{ 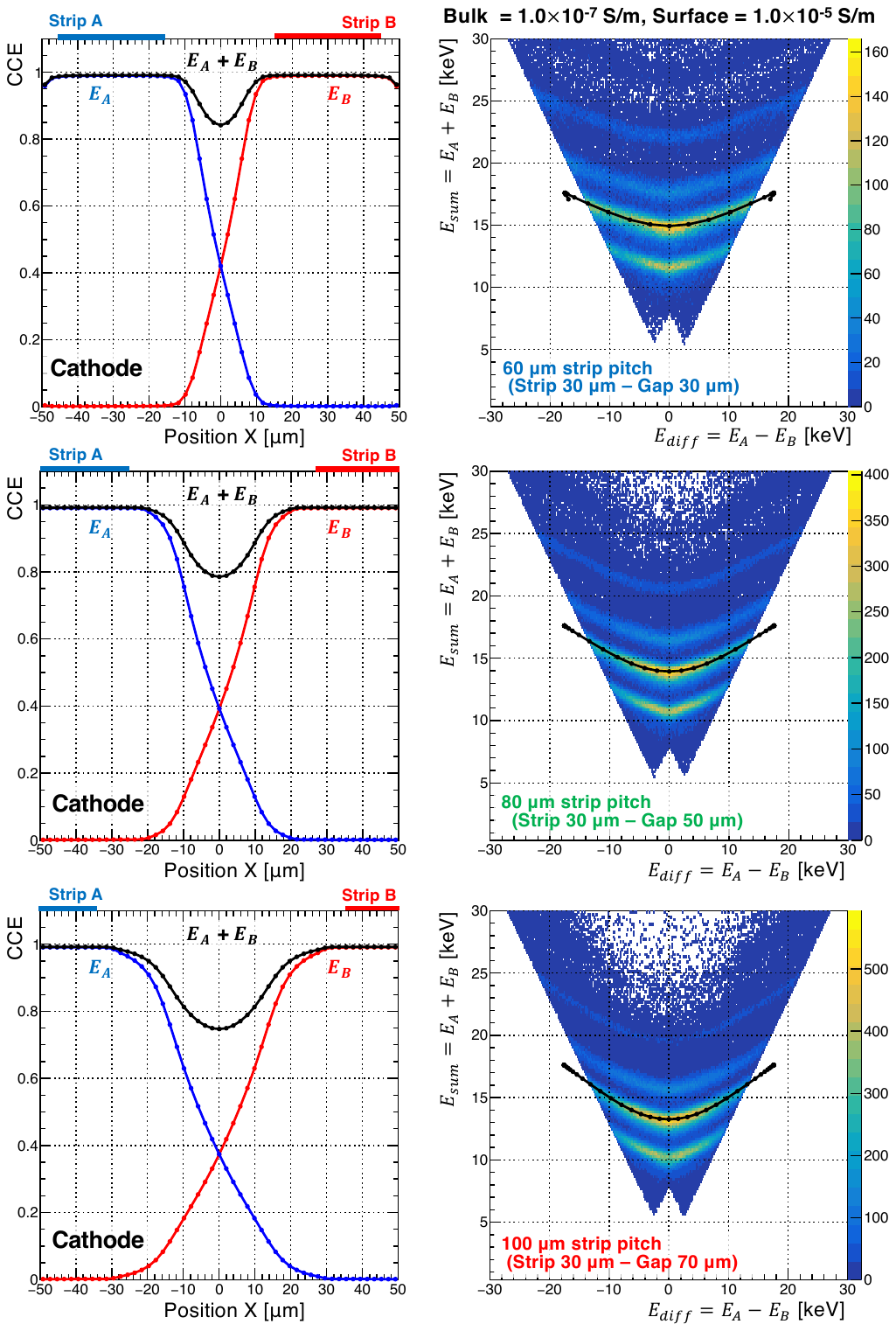}
    \caption{Effects of surface conductivity and gap width on charge loss on the cathode side for each strip/gap configuration. The surface conductivity is fixed at $1 \times 10^{-5}\ \mathrm{S/m}$, two orders of magnitude higher than the bulk conductivity.  (Left) Simulated detected energy in adjacent strips A and B as a function of the incident X-ray position.  (Right) Relationship between the sum and difference of detected energies in adjacent strips (image: experimental results, black dots: simulation).}
    \label{fig:sim_fixcharge}
\end{figure}

\begin{figure}[hp]
    \centering
    \includegraphics[width=0.85\linewidth]{ 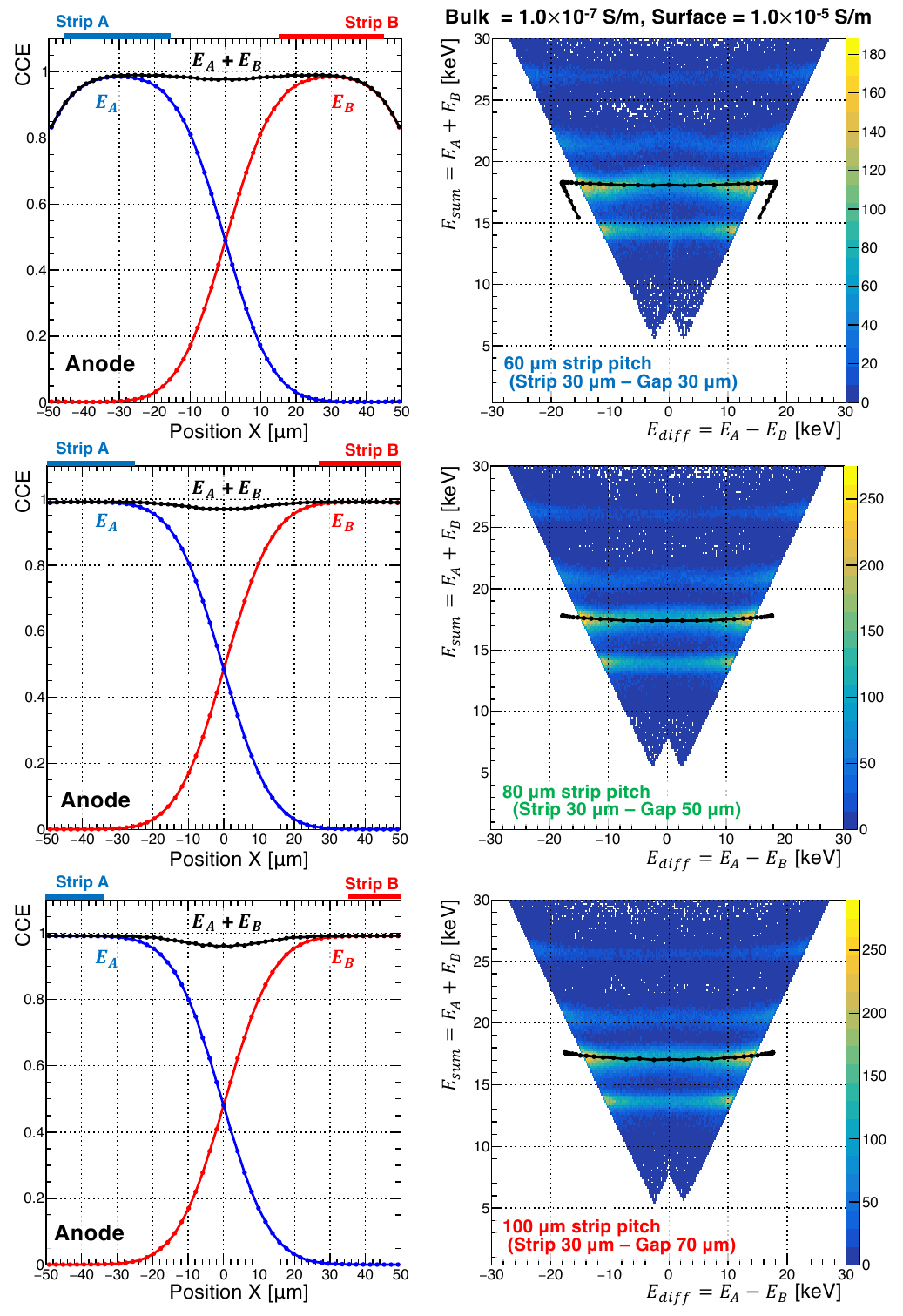}
    \caption{Effects of surface conductivity and gap width on charge loss on the anode side for each strip/gap configuration. The surface conductivity is fixed at $1 \times 10^{-7}\ \mathrm{S/m}$, the same as the bulk conductivity.  (Left) Simulated detected energy in adjacent strips A and B as a function of the incident X-ray position.  (Right) Relationship between the sum and difference of detected energies in adjacent strips (image: experimental results, black dots: simulation).}
    \label{fig:sim_fixcharge_anode}
\end{figure}

Figures~\ref{fig:sim_fixcharge} and~\ref{fig:sim_fixcharge_anode} show the simulated results for the cathode and the anode side.  
In these simulations, we assumed a surface conductive layer of $1 \times 10^{-5}\ \mathrm{S/m}$ on the cathode side for each configuration, while keeping the anode side surface conductivity equal to the bulk value ($1 \times 10^{-7}\ \mathrm{S/m}$).
The charge loss on the cathode side was successfully reproduced across all strip-pitch configurations. 
As the gap width increases, the charge loss becomes larger, and the variation in detected energy relative to the incident X-ray position becomes more gradual.  
The increase in charge loss reflects that a larger fraction of charge carriers drift toward the surface and are not collected efficiently.  
Meanwhile, the more gradual energy variation indicates that charge sharing between adjacent electrodes becomes more prominent with wider gaps.  
These trends are consistent with the experimental results reported in Ref.~\citenum{nagasawa2025imaging}.


Conversely, on the anode side, changing the gap width has little effect on the charge loss, suggesting that the surface layer has a conductivity comparable to that of the bulk and does not significantly distort the electric field.  
The variation of the detected energy relative to the incident X-ray position also remains nearly unchanged with different gap sizes.
Specifically, CCE increases smoothly from zero to one between approximately $-20~\mathrm{\mu m}$ and $+20~\mathrm{\mu m}$ for strip B, showing a similar profile regardless of the gap width.
This suggests that, as the gap widens, the charge sharing region becomes smaller relative to the overall strip pitch.  
This trend is also consistent with the experimental observation of reduced charge sharing events for wider gaps \cite{nagasawa2025imaging}. 

\section{Conclusion}\label{sec:Conclusion}
We have developed a first-principles simulation framework to model the response of wide-gap CdTe double-sided strip detectors (CdTe-DSDs) for the FOXSI-4 and FOXSI-5 solar sounding rocket experiments. 
In our previous work~\cite{nagasawa2025imaging}, we developed phenomenological reconstruction methods based on empirical calibration of the observed detector response, achieving the required sub-keV energy and tens-of-micron position resolution (Section \ref{sec:det_all}).
This work aimed to build a first-principles understanding of charge sharing and charge loss in wide-gap CdTe-DSDs, enabling a more quantitative interpretation of the detector response and supporting the future optimization of detector design and reconstruction algorithms.

By integrating Geant4-based Monte Carlo simulations for energy deposition, finite element method calculations of electric and weighting fields, and charge transport modeling incorporating drift, diffusion, and trapping, we constructed a comprehensive simulation framework that quantitatively reproduces the detector response (Section \ref{sec:sim_all}).
In particular, we found that introducing a surface conductive layer was essential to explain the experimentally observed charge loss on the cathode side (Section \ref{sec:surface}). This surface layer alters the electric field distribution near the surface, directing field lines away from the electrodes and causing charge losses. 
Furthermore, we investigated the dependence of charge loss and charge sharing behavior on gap width (Section \ref{sec:sim_charge_loss}).
On the cathode side, increasing the gap width leads to greater charge loss, as the electric field distortion becomes more pronounced in the wider gap regions.
In addition, the charge sharing region becomes broader with increasing gap width.
In contrast, on the anode side, where the influence of surface conductivity is considered negligible, the charge loss and charge sharing behavior remain largely unchanged across different gap widths.
These trends are consistent with experimental results, demonstrating the validity of our modeling approach.
This simulation framework provides a powerful tool for interpreting FOXSI-4/5 data and for optimizing future CdTe-DSD designs for space-based X-ray imaging spectroscopy. 

\acknowledgments 
This work was supported by JSPS, Japan KAKENHI Grant Number 22J12583.
The authors would like to thank the FOXSI-4 and FOXSI-5 team.
S.N. is supported by FoPM (WISE Program) and JSR Fellowship, The University of Tokyo, Japan, and by the JSPS Overseas Research Fellowship.
This research used computing resources at Kavli IPMU.

\bibliography{report} 
\bibliographystyle{spiebib} 

\end{document}